# Extending OWL-S for the Composition of Web Services Generated With a Legacy Application Wrapper

Bacem Wali

INSERM/INRIA/Univ. Rennes 1, VISAGES U746

Faculté de médecine

Rennes, France

bacem.wali@irisa.fr

Bernard Gibaud

INSERM/INRIA/Univ. Rennes 1, VISAGES U746

Faculté de médecine

Rennes, France

bernard.gibaud@irisa.fr

*Abstract*— **Despite numerous efforts by various developers, web service composition is still a difficult problem to tackle. Lot of progressive research has been made on the development of suitable standards. These researches help to alleviate and overcome some of the web services composition issues. However, the legacy application wrappers generate nonstandard WSDL which hinder the progress. Indeed, in addition to their lack of semantics, WSDLs have sometimes different shapes because they are adapted to circumvent some technical implementation aspect. In this paper, we propose a method for the semi automatic composition of web services in the context of the NeuroLOG project. In this project the reuse of processing tools relies on a legacy application wrapper called jGASW. The paper describes the extensions to OWL-S in order to introduce and enable the composition of web services generated using the jGASW wrapper and also to implement consistency checks regarding these services.**

## I. INTRODUCTION

Web services are a new revolution of software systems. They are considered as self-contained, self-describing, module applications that can be published, located, and invoked through the Web [1][2]. They are designed to be manipulated remotely from a network and they have the capability to invoke each other mutually, which raises the issue of their interoperability. Companies implement web services according to their application domain and display them through the web. Consequently, the number of heterogeneous web services is increasing, whose interoperability is severely hampered by its pervasive heterogeneity, inherent to independently developed services. For example: The use of new messaging protocols involves changing WSDL formats according to the domain specific applications and implementation needs. Therefore, when we deviate from standard cases to specific ones, composition of web service becomes a challenging problem that was addressed by many researchers and engineers in the recent years [3].

Different initiatives have been proposed to facilitate the reuse of web services, leading to new languages, protocols and frameworks. For example, UDDI [4] (Universal Description, Discovery and integration), SOA (Service Oriented architecture) [5], BPEL4WS (Business Process Execution Language for Web Service) [6], SOAP [7] (Simple Object Access Protocol) and WSDL (Web Services Description Language) [8] are standards for service discovery, description, and messaging protocols [1][9]. Those specifications provide a means to syntactically describe a web service. However, they do not deal with semantic web service description and semantic web service composition.

Semantic web service is a concept that brings semantics to the aforementioned standards. By adding semantics we can make web services machine understandable and use-apparent form [10]. By adding semantic markup to a web service we can make two aspects of its functionality explicit. First, semantic annotations can define what the service actually does, and second, they can describe its behavioral aspect (i.e., how the service works, the chaining that can be performed according to the sent and received messages).

Several languages have emerged, to add semantic description features to the web services standards. For example, DAML-S [11] (Darpa Agent Markup Language for services) is a revision of DAML+OIL [12] and based on OWL [13] (Ontology Web Language) WSMO [14] (Web Service Modeling Ontology) and OWL-S (Ontology Web Language for Services) [15]. OWL-S is an ontology represented in OWL which aims at applying reasoning capabilities to the functionality, behavior, and execution of web services. OWL-S defines a model with three layers: a **service profile** describing the service's basic functionalities (function and characteristics, etc), a **service model** describing how the service works including data and control constructs flow, and a **service grounding**, describing how to access the service, by grounding its functional elements (input, outputs and operations) in a way consistent with the WSDL's concept of binding.

We worked in the context of the NeuroLOG [16] project which aims to share medical resources (brain images and image processing tools) [17]. Image processing tools are wrapped as web services using a software package called jGASW [18]. JGASW is a framework for wrapping legacy scientific applications as web services enabling their execution in a SOA environment.

To allow the sharing of neuro-imaging resources, the OntoNeuroLOG [19] ontology was designed. It provides common semantics for information sharing throughout the NeuroLOG system and allows the sharing of neuro-imaging resources provided by collaborating actors in the field of neuro-imaging research**.** The term resources cover both neuro-imaging data (such as images) as well as image processing tools (registration, de-noising, and segmentation).

Through the semantic we tend to share the functionalities of images and the functionalities of image processing tools to enable more expressivity from a functional viewpoint.

The goal of our work is to facilitate the sharing, reuse, and invocation for the user of image processing tools wrapped as web services with jGASW and deployed in the site server within the NeuroLOG framework. To this end, we chose to add semantics to jGASW services, to facilitate workflow composition and automate some consistency controls regarding their usage. Our assumption is that this may increase the usability of such tools by people that were not involved in their development. This by addressing some technical aspects that hinder the composition process with OWL-S API and implementing some consistency controls. Such consistency control tends to: (i) ensure interoperability and composition by checking the compatibility between outputs and inputs of web services; (ii) check of the compatibility between the inputs provided by the users and the semantic inputs definition of the service; (iii) check the consistency between the functionality of the image processing tool, like registration or de-noising, and their declared inputs/outputs, with respect to the formal definition of such conceptual actions, modeled in the OntoNeuroLOG ontology.

We used the OWL-S ontology to benefit from its web service description model and its large expressivity in terms of parameters description and behavioral aspect of flows. However, we had to deal with some technical issues regarding jGASW WSDLs which let us extending OWL-S to enable its use in our NeuroLOG framework.

In this paper, we advance the state of the art by (1) specifying an extension of the OWL-S specification to make it adapted to our jGASW framework context without changing the basic structure of the WSDLs, (2) adding some reasoning capabilities to perform consistency checks regarding the usage of our annotated web services. The following of the paper is organized as follows: Section II provides more details about the difficulties related to the WSDLs of service generated by the jGASW tool, together with OWL-S semantic descriptions of services. Section III presents the solutions that we found based on a specific extension of OWL-S that addresses the problem and some associated reasoning mechanisms that validate the services capabilities, consistent with our domain ontology OntoNeuroLOG. Section IV details how the implementation was done to extend OWL-S and solve the problem and describes some technical issues that we tackle. Section V discusses our contribution and situates it in the wider context of semantic workflows, and finally, Section VI opens other perspectives for future work.

## II. BACKGROUND

To address the issue of web service composition using OWL-S and jGASW we need to describe both more closely. So, we first describe WSDL files generated automatically by jGASW, then we study the automatic generation of semantic descriptions with OWL-S and we analyze the mismatch between the two and we study how to address it.

First, we explain how the jGASW [11] framework works: an application GUI allows the user to upload the image processing tools (shell program) and add inputs, outputs arguments and libraries. According to an XML schema and values set by the user, an XML description is generated *jGASW descriptor*. The generation of the web service consists in transforming the *jGASW descriptor* into a web service interface by generating the WSDL together with an XSD schema (XML Schema Definition). The XSD schema details the inputs and the outputs of every WSDL operation. In fact, all services have WSDL files with the same content, and identical operations but always have only one input and one output and one output as an exception. In contrast, inputs and outputs are described differently in the XSD schema according to each service. For example, in the Figure 1, column 1 illustrates the description of a WSDL operation named *local* that is composed of *tns:local* as input, *tns:localResponse* as output, and *tns:SOAPException* as fault message (generated if the execution of the service failed). Column 2 details input *tns:local* by defining it as a **complexType containing two xs:element** (i) (simpleinput1 and simpleinput2: input files) and details output *tns:localResponse* by defining it as a **complexType typed as another complexType** (ii) (jigsawOutputTest2111: generated automatically). This complexType contains four **xs:element** representing the different output files (stdout, stderr, simpleoutput1, simpleoutput2).

Column 3 shows SOAP envelopes (call/request). At execution time, jGASW prepares the SOAP envelope, invokes the service and gets back the result according to the sequences described in the complex type jigsawOutputTest2111, but the name of the envelope is *ns1:localResult*. Thus, the jigaswOutputTest211 is not considered here.

JGASW wraps executables into web services, and produces at least two standard files (std.out: for standard output for any shell output, std.err: error message generated (if execution fails), and other files resulting of the service execution such as image files (e.g. ex2output1.nii, ex2output2.nii).

Column 4 in the figure 1 shows the OWL-S description generated automatically from the WSDL description provided by the jGASW service (using the WSDL2OWLS [20] converter). Green arrows show the grounding of each individual input element whereas the red arrow shows one unique grounding which is *localResut*. In fact, here, we lost the information that this output contains four files rather one only. Thus, composition of jGASW services is not possible.

So, the principal issue concerns the outputs definition, which is not understandable due to their complex schema.

Figure 1: Automatic grounding of a jGASW service using the WSDL2OWLS API

This figure shows the main problem faced when we try to automatically generate the semantic description of a jGASW service; Columns 3 and 4 show how inputs are grounded individually whereas outputs are grounded as a single box "locaResult". Therefore, OWL-S Process and Profile of this service contain one single output according to the obtained grounding. Why doesn't OWL-S understand different outputs? Because they belong to a complex type "jigsawOutputTest2111" showed on column 2. Such output types generated automatically by the jGASW software can be even more complex than that. Actually, they can contain multiple nestings of complexType.

**Motivation:** First, we have shown that WSDL files are only partially understood by OWL-S API. Second, we need *control construct* to design and execute our image processing workflows. Third, OWL-S is a well defined language for web services composition that offers many functionalities that are not handled by other languages. Moreover it is submitted to the W3C , something important in our context of collaborative research in neuro-imaging. This choice of OWL-S is also consistent with our use of the OWL-Lite OntoNeuroLOG ontology as domain ontology. Indeed, OWL-S is an OWL-DL ontology and it is more expedient to use ontologies that are closer to each other in term of semantic capability and reasoning (i) both are based in OWL and this facilitate the use the same kind of reasoner (ii) if we use WSMO for example we should translate OntoNeuroLOG in WSML. Finally, OWL-S provides the suitable expressivity for representing web service semantics and fits nicely our application's requirements; in fact we cannot modify the structure of WSDL files, nor the XML description of inputs/outputs, since they are intrinsic to the jGASW middleware (otherwise invocation would not work properly).

**Approach:** First we extend the OWL-S Profile to be adapted to OntoNeuroLOG ontology. Second, we extend OWL-S Process to enable the description of jGASW services and finally, we add a software layer that implements reasoning services ensuring various consistency checks based on knowledge imbedded in the OntoNeuroLOG ontology.

## III.  METHOD

The OntoNeuroLOG ontology describes the different kinds of brain images in reference to the Dataset taxonomy and the functionality of services in reference to the Data processing taxonomy, each of these classes having its specific characteristics defined using DL axioms.

### A.  Extending OWL-S

OWL-S is a particular OWL ontology. It allows the semi-automatic composition of Web services. It is composed of three layers. The Service Profile allows the description, publication, and discovery of services. It is used by providers to publish their services and by users to specify their needs. The Service Model is used to compose services. It allows modeling services as processes. Three types of processes exist: atomic processes (AtomicProcess), simple (SimpleProcess) and composite (CompositeProcess). AtomicProcess represents the finest level of action that the service may perform. Composite Process are decomposable into other processes thus their concatenation can be specified using a set of control structures such as Sequence, Split, If-Then-Else, etc. A SimpleProcess is used to provide a view of an atomic process or a simplified representation of a composite process.

The Service Grounding describes how to access the service and provides the mapping between semantic inputs, outputs, message formats, and physical addresses. The purpose of this mapping is to enable the translation of semantic inputs generated by a service consumer into the appropriate WSDL messages for transmission to the service provider, and the translation of service output messages back into appropriate semantic descriptions (i.e., OWL descriptions) for interpretation by the service consumer.

#### 1)  Extending the OWL-S Process model

Our extension aims at decomposing the output grounded as a single box (localResult) with OWL-S (described in Figure 1) into its different elements (e.g., stderr, stdout, simpleoutput1, simpleoutput2). To overcome this problem, some classes and data/object properties are added to the OWL-S process model:

- a NlogParameter class to denote parameters that are embedded in a parameter of such a composite nature (e.g., stderr, stdout, simpleoutput1, simpleoutput2); it is defined as:

```
<owl:Class rdf:about="#NlogParameter">
    <rdfs:subClassOf rdf:resource="#Parameter"/>
    <owl:disjointWith rdf:resource="#Output"/>
</owl:Class>
```

- a nlogExpandsTo object property, associating a parameter of a composite nature to its essential elements according to the XSD schema of the service;

```
<owl:ObjectProperty rdf:about="#nlogExpandsTo">
    <rdfs:domain rdf:resource="#Output"/>
    <rdfs:range rdf:resource="#NlogParameter"/>
</owl:ObjectProperty>
```

- a hasID data property, denoting the markup within the string result after the service invocation

```
<owl:DatatypeProperty rdf:about="#hasID">
    <rdfs:domain rdf:resource="#NlogParameter"/>
    <rdfs:range rdf:resource="&xsd;anyURI"/>
</owl:DatatypeProperty>
```

- a hasLabel data property, denoting a non-functional property providing an informal description of the parameter

```
<owl:DatatypeProperty rdf:about="#hasLabel">
    <rdfs:domain rdf:resource="#NlogParameter"/>
    <rdfs:range rdf:resource="&xsd;anyURI"/>
</owl:DatatypeProperty>
```

Figure 2 provides an illustrative example of the use of the previous extensions: we present one extension of the output of jGASW service?

```
<process:Output rdf:about="&tool;#Atomic_Process_Test2_output1">
    <process:nlogExpandsTo><process:NlogParameter
    rdf:ID="Atomic_Process_Test2_output1_simpleoutput1">
    <process:hasLabel rdf:datatype=
    "http://www.w3.org/2001/XMLSchema#anyURI">
            This is the extension of the first parameter simpleoutput1
    </process:hasLabel><process:hasID rdf:datatype=
    "http://www.w3.org/2001/XMLSchema#anyURI">
            simpleoutput1
    </process:hasID> <process:parameterType rdf:datatype=
    "http://www.w3.org/2001/XMLSchema#anyURI">
    http://localhost/dataset-owl-lite.owl#T1-weighted-MR-dataset
    </process:parameterType>
    </process:NlogParameter></process:nlogExpandsTo> </process:Output>
```

Figure 2: Enrich the output in its essential parameters

To detect the value of "simpleoutput1" argument first we select its ID using the hasID property and second we parse the string result (SOAP envelop) using the retrieved ID corresponding to the markup of "simpleoutput1". The data type of the NlogParameter is given by the data property *Process:parameterType*.

### 2) Extending the OWL-S Profile model

As described before, the OWL-S Profile gives information about the capabilities and the behavior of the service. We enrich it by adding a reference to the equivalent data processing class using refers-to, an object property that belongs to OntoNeuroLOG.

```
<owl:ObjectProperty rdf:about="&iec;refers-to">
    <rdfs:domain rdf:resource="#Profile"/>
    <rdfs:range rdf:resource="&data-processing-owl-lite;data-processing"/>
</owl:ObjectProperty>
```

### 3) Web services composition

As mentioned in the background section we cannot modify the WSDL otherwise invocation would no longer work, in consequence we could not extend the Grounding sub-ontology. The extension of the Process Model is enough to allow jGASW services composition. Once OWL-S Outputs have been related to corresponding *NlogParameters* according to the XSD Schema derived from jGASW processing, we were able to compose jGASW services. To this end, we introduced another object property, *links*, that binds any OWL-S Parameter to another (also suitable to NlogParameter since they are a subClassOf Parameter):

```
<owl:ObjectProperty rdf:about="#links">
    <rdfs:domain rdf:resource="#Parameter"/>
    <rdfs:range rdf:resource="#Parameter"/>
</owl:ObjectProperty>
```

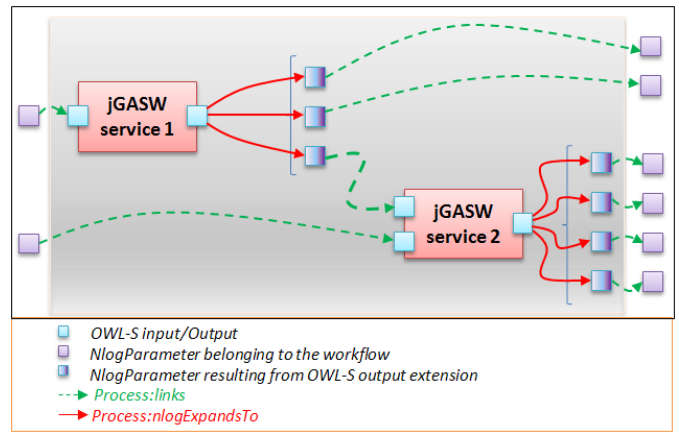

Figure 3 : How to link NlogParameters with OWL-S parameters and workflow parameters in case of workflow composition

In this illustrative example, we compose two jGASW services: the first one (service 1) has one input and one output. The output is composed of three outputs according to its XSD Schema, and the second (service 2) has 2 inputs and one output. The output is composed of four outputs according to its XSD schema. The profile of the service embedding the whole workflow has two inputs linked respectively to jGASW service 1 and jGASW service 2 and six outputs coming from both jGASW services. One internal parameter only is transmitted from service 1 to service 2.

### B. Some reasoning mechanisms

#### 1) Compatibility check between dataset processing and the OWL-S Profile

This service allows users to ensure that the definition of the profile is compatible with the data processing class selected by the user at annotation time.

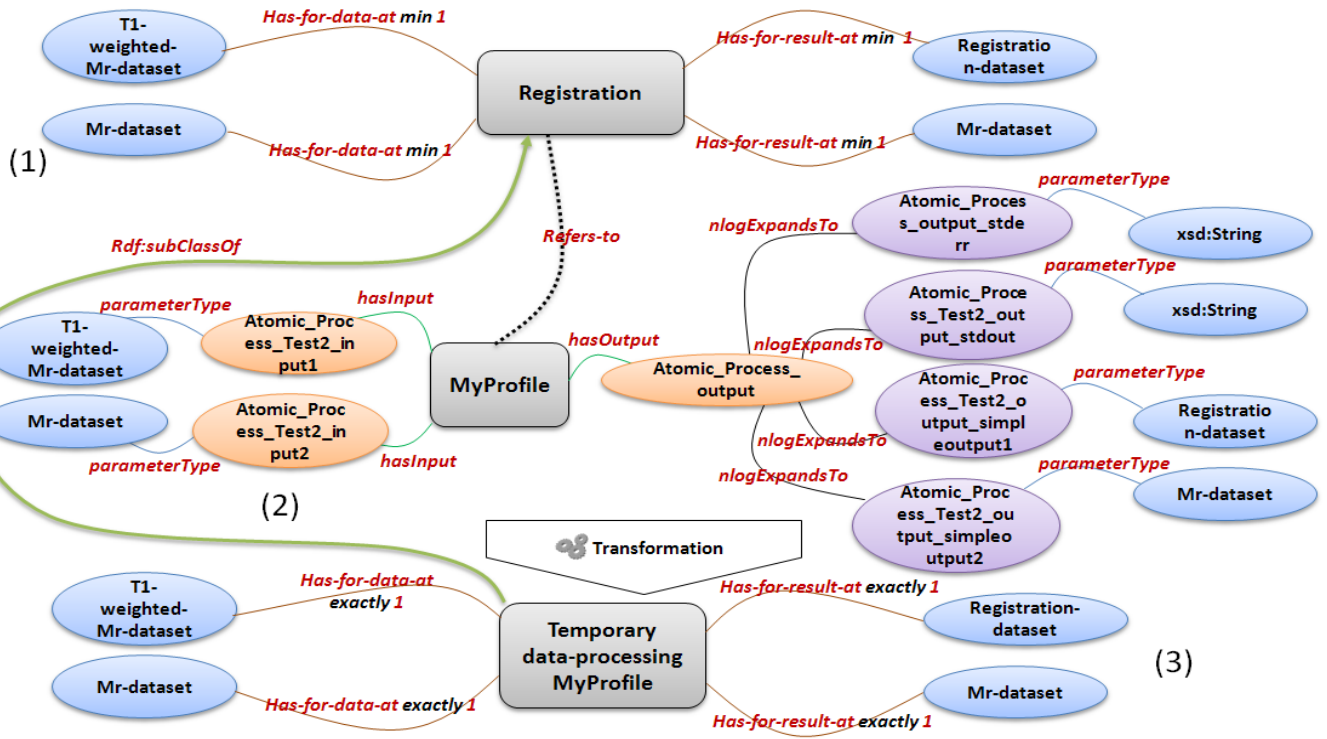

Figure 4: Transformation of Profile to data processing class

(1) Represents the description of Registration data processing, (2) represents the semantic description of the registration tool according to enriched OWL-S that should do registration if invoked and (3) shows the transformation of the profile into data processing.

The algorithm is the following: first we create a temporary class *tmp_Profile_data-processing* class relatively to the current operation, and then we translate relations between profile, inputs, and outputs into axioms and we add profile them to the *tmp_Profile_data-processing* class. Then, for every relation hasInput/hasOutput we count the number of inputs grouped by dataset class to determine the cardinality of the corresponding axiom; for example: *Process:hasInput* **i1** *Process:parameterType* Mr-dataset and *Process:hasInput* **i2** *Process:parameterType* Mr-dataset would lead to a cardinality of 2 concerning Mr-dataset (Mr-dataset denotes a magnetic resonance image dataset). The third step consists in selecting the appropriate object property for the construction of the axiom (e.g. *Process:hasInput* substituted by *has-for-data-at* and *Process:hasOutput* substituted by *has-for-result-at*. The result of the two first steps is: (*Process:hasInput* **i1** *Process:parameterType* **Mr-dataset** and *Process:hasInput* **i2** *Process:parameterType* **Mr-dataset**) ➔ (*has-for-data-at* **exactly 2** Mr-dataset). The third step consists of adding those axioms to the *tmp_Profile_data-processing* class.

The last step is to add the new tmp_Profile_data-processing class with axioms added above as subclass of the class referred by the Profile "MyProfile" and selected by the user, in our example (tmp_Profile_data-processing subclassOf Registration), and then, classify and check consistency. If the ontology is consistent then the annotation is considered valid. Semantically, the functionality of the tool is agreed, i.e., the has-for-data-at/has-for-result-at object properties are consistent with respective inputs/outputs specified in the corresponding data processing class in the OntoNeuroLOG ontology. Figure 4 show an illustrative example of the algorithm.

*2) Compatibility check between outputs and inputs in a workflow*

This service is applied when a user builds a new workflow. The processing aims at ensuring for every link between NlogParameter and Input that corresponding types are compatible. So we distinguish three cases:

• Identical data types: the output and the input have exactly the same type. Compatibility is validated and composition is accepted.

• Link to a more specific data type: the output is more general than the input of the next service, so non-compatibility.

• Link to a more general data type: the output is more specific than the input of the next service. The first service will always return results that are semantically compatible with the next service input. Compatibility is validated and composition is accepted.

N.B. workflow is valid if Parameters have the same Type or source is subsumed by target according to the dataset ontology.

*3) Compatibility check between values and inputs at invocation time*

This service is called when a web service is invoked. It checks whether the actual instances selected by the user (e.g. a Dataset) and assigned to the values actually meet the constraints specified in the semantic annotations of the service. In practice, the semantic service checks whether the class (or the type) of this instance is subsumed by the class type of the input.

## IV. IMPLEMENTATION

The semantic annotation of jGASW services is generated automatically using the WSDL2OWLS API. Enrichment of semantic annotation is done using the OWL-S 1.2 specification and the OWLS API 3.0.
The semantic annotation of workflow services is generated using the OWL-S 1.2 specification and the OWLS API 3.0.

The consistency check between the profile and the data processing class is implemented using the OWL API, the OWL-S API 3.0 and the HermiT Reasoner.

The web services invocations use the OWL-S API but results and composition issues use the semantic search engine CORESE [21] together with the OWL-S API. CORESE is used to select the functional properties (extensions, linked parameters, identifiers …) of OWL-S outputs by querying the triple store containing the semantic annotations of the services. We add here an illustrative example of a workflow composed of two services (Figure 5). First, we prepare the SOAP envelope to invoke the first jGASW service: green markup shows the WSDL operation input (tns:local) and blue markup indicates the concrete input that the service will use.

```
<soapenv:Envelope><soapenv:Body> <local xmlns="http://i3s.
cnrs.fr/jigsaw"><simpleinput  xsi:type="xsd:string"> http
//i3s.cnrs.fr/jigsaw">http://localhost/test1.nii</simpleinput>
</local></soapenv:Body></soapenv:Envelope>
```

The next section shows the output of the service after invocation: green markup shows the output of the WSDL operation (tns:localResult). It wraps three blue markups that show three files generated by the service.

```
<ns1:localResult        xmlns:ns1="http://i3s.cnrs.fr/jigsaw">
<stderr>http://localhost:80/~bwali/Test1_1321350928548-9787/std.err
</stderr>    <stdout>http://localhost:80/~bwali/Test1_1321350928548-
9787/std.out</stdout><simpleoutput>http://localhost:80/~bwali/Test1
_1321350928548-9787/testoutput.nii</simpleoutput>
</ns1:localResult>
```

The stderr and stdout are workflow outputs whereas simpleoutput should be transmitted to the second jGASW service. With CORESE we query the triple store to retrieve the nlogParameters to which the output (localResult) is extended. Then, for every nlogParameter retrieved, we find the ID (the markup to extract it from the localResult), its link to another parameter (parameter passing), and its data type. The parameter that we extract is simpleoutput. It should be transmitted to ex002_input2 second input of the second jGASW service.

**Query:** aims at identifying the different outputs that tns:localResult (corresponding semantically to *ex001_output1*) is expanded to:

```
PREFIX p1: <http://localhost/kb/Test1_2.owl#>
PREFIX p2: <http://localhost/Process.owl#>
Select ?nlogParameter  ?link  ?id  ?type  where {
p1: ex001_output1  p2:nlogExpandsTo  ?nlogParameter
?nlogParameter    p2:links          ?link
?nlogParameter    p2:hasID          ?id
```

```
?nlogParameter     p2:parameterType    ?type }
```
**Query Results:**

<span style="color:red">?nlogParameter</span>  http://localhost/kb/extension-Test1_2.owl#ex001
_simpleoutput

<span style="color:green">?link</span>  http://localhost/kb/Test1_2.owl#ex002_input2

<span style="color:blue">?id</span>  simpleoutput

<span style="color:purple">?type</span>  http://localhost/dataset-owl-lite.owl#T1-weighted-MR-dataset

<span style="color:red">?nlogParameter</span>  http://localhost/kb/extension-Test1_2.owl#ex001
_stdout

<span style="color:green">?link</span>  http://localhost/kb/extension-Test1_2.owl#WF_stdout

<span style="color:blue">?id</span>  stdout

<span style="color:purple">?type</span>  http://www.w3.org/2001/XMLSchema#string

<span style="color:red">?nlogParameter</span>  http://localhost/kb/extension-Test1_2.owl# ex001_
stderr

<span style="color:green">?link</span>  http://localhost/kb/extension-Test1_2.owl#WF_stderr

<span style="color:blue">?id</span>  stderr

<span style="color:purple">?type</span>  http://www.w3.org/2001/XMLSchema#string

The output ex001_output1 is expanded to three nlogParameters as seen in the Figure 5 (ex001_stdout, ex001_stderr, ex001_simpleoutput) corresponding respectively to (stdout, stderr, simpleoutput) in the query results (?id fields). Those ID are the markups used in the localResult. The query results show that both ex001_stdout, ex001_stderr are linked to workflow outputs (WF_stdout1, WF_std_err1) as showed in the Figure 5. The query results show that the parameter ex001_simpleoutput is linked to the parameter ex002_input2.

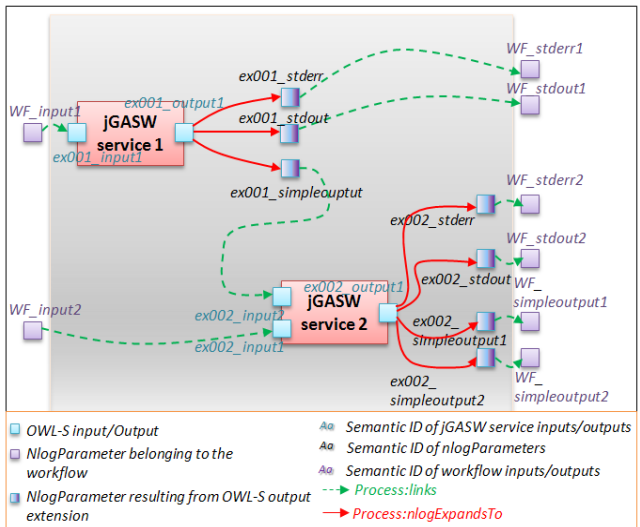

Figure 5: semantic annotation of workflow using the OWL-S Process layer and the extension described in this work

Thus it should be passed to the second jGASW service. To this end, the value of *ex001_simpleoutput* is extracted using the jGASW engine by giving the ID already selected by the query. The result is:<span style="color:red">http://localhost:80/-
~bwali/Test1_1321350928548-9787/testoutput.nii.</span> A new SOAP envelope containing two inputs (as Figure 5 shows) is prepared to invoke the second jGASW service.

```
<soapenv:Envelope     <soapenv:Body>><local  xmlns="http://
i3s.cnrs.fr/jigsaw"><simpleinput1     xsi:type="  xsd:string"xmlns=
"http://i3s.cnrs.fr/jigsaw"> http://localhost/test4.nii </simpleinput1>
<simpleinput2  xsi:type="xsd:string   xmlns="http://i3s.cnrs.fr/jigsaw">
http://localhost:80/~bwali/Test1_1321350928548-9787/testoutput.nii
</simpleinput2> </local></soapenv:Body></soapenv:Envelope>
```

The simpleinput1 is the file selected by the user for the workflow execution (corresponding semantic id is *WF_input2*). This parameter is passed to *ex002_input2*. The simpleinput2 gets the file extracted from localResult. i.e. result of the execution of first jGASW service invoked as described above. The other parameters (stderr, stdout, simpleoutput1 and simpleoutput2 of jGASW service2) are transmitted to the workflow outputs.

## V. DISCUSSION

Several semantic languages and frameworks have been proposed based on W3C web service languages to support web service composition. However, web service composition is hampered by the heterogeneity of web services. Our work is an extension of OWL-S at the concrete service level to address the issue of jGASW web services composition.

We relied on OWL-S because it is a well defined Ontology [21] based on manifold earlier solutions and it is currently submitted in the W3C. It is also a semantic framework that provides more complete specifications than any other alternative solutions. It is represented in OWL which is a standardized language and exploits its reasoning capability [22]. Thus, it enables us to leverage our domain ontology in reasoning aiming at performing various consistency checks regarding the use of our services. OWL-S is a multi-layered language thus, it is easy to handle. In our contribution, extending the Profile layer and the Process layer leverages this characteristic. OWL-S differs from other specifications by providing conditions, effects, sequences and control constructs. We reused conditions and effects definitions to verify the consistency of service compositions and control construct specifying the behavioral aspect [23] of composed jGASW services. The OWL-S Service Grounding is conceived to be adapted for grounding any kind of service. Unfortunately, our WSDL files are really specific and cannot be grounded entirely. Getting service output as a unique box and as a string format actually hampers generating the grounding automatically and therefore the semantic description. Nevertheless, OWL-S is still the nearest solution and its adoption and extension allowed overcoming the problem.

WSDL-S [24] and SAWSDL [25] define how to add semantic annotations to WSDL specifications. In fact, they let WSDL components refer to semantic concepts via the **ModelReference** attribute, added to WSDL elements to assign one or more semantic concepts, via the **schemaMapping** property to map complex types and elements with a semantic model, via **Precondition** and **effect** for service discovery, via **serviceCategory** to help in case of service advertisement. In contrast to OWL-S they externalized domain application and let the reasoning mechanisms free. Grounding should be interpreted manually and service composition is not explicit. They do not deal with context of execution, behavior aspect and therefore, the reasoning aspect is really neglected, so we preferred use a more sophisticated and developed language for reasoning mechanisms.

Web service composition is still a complex task [1][26][27]. Numerous surveys on web service composition present an overview of methods that deal with web service composition. Based on a large background, Dustdar and Schreiner [27] discussed the need of web service composition and related issues. They outline the importance of the *context* in web service composition. The *context* should be formatted in some customized and personalized manner for relevant use by the next service. In our work we had to face the same requirements regarding the composition problem. The enrichment of OWL-S aims to format outputs in order to make them adequate for the next service that will be invoked. Enabling jGASW services composition is the added value of this enrichment and key factor of our work. It enabled us to add algorithms to check consistency

Rao and Su [1] investigated automated web service composition and propose an abstract framework for automatic service composition. They discuss abstract process model and business workflow involving the impact of *heterogeneity* of web services sources. We conclude that web service composition becomes more difficult if ever we deviate from the standard cases to specific cases. For example, automatic selection, matching, and composition work well while using standards. It is against hindered if we are out of standards. In our work, jGASW WSDL files are different from standard WSDL files. They differ by their XML schema, thus, they are heterogeneous compared to standard ones. This shows nevertheless, the dependency of semantics model on thin technical details and with the manner how to access services.

Without OWL-S the composition of jGASW service is not possible. In fact the form of SOAP envelop of the result not allows the chaining of web services. If we would like to compose jGASW services without semantics we should add interoperability within the jGASW engine. The first benefit from extension and use of OWL-S facilitate this task by enable the composition process. The second benefit from OWL-S is the multilayered structure that it has. In fact with the ServiceProfile it enables us to add semantic verification according to the neuro-imaging expectations. This is shown throw the implementation of the validation algorithm.

Casati et al [28] uses the notion of process template to model composite services and composers need to browse the process library to search for process templates of interest [27]. Rao et al [1] and Dustdar et al [27] distinguish in workflow composition static and dynamic workflow generation, static defines the business tasks and dynamic linking the concrete e-services. Both help for monitoring e-services. OWL-S does not provide explicit support for monitoring and errors handling [29]. OWL-S service profile is just a service categorization and still lacks semantics. In our work we add some semantics to augment workflow monitoring. For example while users compose their workflows our consistency checking algorithm verifies that the service profile and the related data processing are consistent, which ensures that the chaining of the service can make sense from the point of view of processing.

Cardoso and ShethIn [30] try to overcome e-workflow composition problems by making services interoperable. They use a multidimensional approach based on ontology mediation. Medjahed et al [26] address the interoperability issue by using composability rules. Currently, this task must be performed by a human who might use a search engine to find a service, and connect the service manually. However, a couple of verification algorithms were implemented within our application framework using OWL-S markup of services, and the necessary information from the OntoNeuroLOG ontology. At this stage, our work is still basic, the automatic discovery and mediation process are not well handled. Indeed, this process requires further development to overcome the heterogeneity of semantic web services using mediation. Especially WSMO, that uses the mechanism of mediation between semantic services coming from different heterogeneous frameworks. In our work the semi-automatic composition does not need mediation, however it needs a semantic validation through a reasoning aspect implementing verification of the consistency of the flows.

Gannod et al. [31] the authors present a generic approach to ground services with OWL-S. Users can ground automatically or manually the service to its description. Although it is a generic approach this kind of grounding does not meet our needs. In fact, it considers that every end point (WSDL or others) defines the outputs individually. However in our case the outputs are embedded in the unique box and are not explicit for the WSDL API and are understandable only by our jGASW Engine. Semantically, this editor considers that service grounding and service model are two distinct layers. In our work, we no longer keep those two layers separate, which is a limitation of our solution. In fact the process:hasID data property that was added is the unique way to access to the WSDL elements (input/output) as explained in the implementation section . We are required to do that because the way jGASW gets back the result obliges us to have a link to the parameter in the process specification. Otherwise, if we try to extend service grounding, invocation would no longer work.

Web services differ in form, technique or design point of view. Application wrappers provide outputs and inputs in different forms due to the functional requirements of the application domains. In this case, semantic solutions are not enough. The extension that we proposed is adequate for every kind of service so we augment the flexibility of web service development. Even the service has different technical details, the proposed idea, when reused in another context, is still valid and address both technical and semantic problems.

## VI. CONCLUSION AND FUTURE WORK

In this paper, we introduced a method to extend the OWL-S specification to cope with jGASW web services description. We succeeded to address the problem of semantic web services composition and to add some semantic validation and verification mechanisms. This solution addresses several issues concerning the web services composition in the neuro-imaging domain, but it is not sufficiently tested by NeuroLOG users. For us to assess its added value from an end user point of view, moreover, automatic composition still needed.

The next step of this work tends to, ensure and validate this work by adding serious test through the neuro-imaging framework and developing an algorithm for automatic selection and composition of jGASW web services and OWL-S workflows. We are trying to add reasoning capability over the description of data processing and the validation with profile algorithm.

### ACKNOWLEDGMENT

The NeuroLOG project was funded by the French National Agency for Research (ANR-06-TLOG-024). The authors gratefully acknowledge all contributors, including our clinical colleagues providing the image data exploited in the NeuroLOG test-bed.